\documentclass[11pt,twoside,letterpaper]{article} 
\usepackage{times,fancyhdr}
\usepackage[dvips]{graphicx}
\makeatletter 
\def\cleardoublepage{\clearpage\if@twoside \ifodd\c@page\else%
    \hbox{}%
    \thispagestyle{empty}%
    \newpage%
    \if@twocolumn\hbox{}\newpage\fi\fi\fi} 
\makeatother 
\def\figurename{Figure}
\makeatletter
\renewcommand{\fnum@figure}[1]{\figurename~\thefigure.}
\makeatother
\def\tablename{Table}
\makeatletter
\renewcommand{\fnum@table}[1]{\tablename~\thetable.}
\makeatother
\begin{document}
\title{
{\begin{flushleft}
\vskip 0.45in
{\normalsize\bfseries\textit{Chapter~1}}
\end{flushleft}
\vskip 0.45in
%
%
%
%
\bfseries\scshape Pairing correlations and thermodynamic properties of inner crust matter}}
\author{\bfseries\itshape  J\'er\^ome Margueron$^1$\thanks{E-mail jerome.margueron@ipno.in2p3.fr}
and Nicolae Sandulescu$^2$\thanks{E-mail sandulescu@theory.nipne.ro}
\\
$^1$ Institut de Physique Nucl\'eaire, IN2P3-CNRS and Universit\'e Paris-Sud, \\F-91406 Orsay CEDEX, France\\
$^2$ Institute of Physics and Nuclear Engineering, 76900 Bucharest, Romania}
\date{\today}
\maketitle

\thispagestyle{empty}
\setcounter{page}{1}
\thispagestyle{fancy}
\fancyhead{}
\fancyhead[L]{In: Neutron Star Crust \\ 
Editors: C.A. Bertulani and J. Piekarewicz, pp. {\thepage-\pageref{lastpage-01}}} 
\fancyhead[R]{ISBN 0000000000  \\
\copyright~2012 Nova Science Publishers, Inc.}
\fancyfoot{}
\renewcommand{\headrulewidth}{0pt}
%
%
\vspace{.08in} \noindent \textbf{Keywords:} Neutron star crust, nuclear superfluidity.
%
\pagestyle{fancy}
\fancyhead{}
\fancyhead[EC]{J. Margueron \& N. Sandulescu}
\fancyhead[EL,OR]{\thepage}
\fancyhead[OC]{Pairing correlations and thermodynamic properties of inner crust matter}
\fancyfoot{}
\renewcommand\headrulewidth{0.5pt} 
%
\begin{abstract}
In this review paper we discuss the effects of pairing correlations on inner 
crust matter in the density region where nuclear clusters are supposed to
coexist with non-localised  neutrons. The pairing correlations are treated
in the framework of the finite temperature
Hartree-Fock-Bogoliubov approach and using zero range nuclear forces. 
After a short introduction and presentation of the formalism
we discuss how the pairing correlations
affect the structure of the inner crust matter, i.e., the proton to neutron
ratio and the size of Wigner-Seitz cells. Then we show how the pairing 
correlations influence, though the specific heat of neutrons, the thermalization
of the crust in the case of a rapid cooling scenario.
\end{abstract}

%
%
%
%
\newpage
\section{Introduction}
\label{sec:introduction}

The superfluid properties of the inner crust of neutron stars have been 
considered long ago in connection to the  large relaxation times which
follow the giant glitches. Thus, according to the present 
models, the glitches are supposed to be generated by the unpinning of
the superfluid vortex lines from the nuclear clusters immersed in the
inner crust of neutron stars ~\cite{anderson1975,pines}.
Later on the superfluidity of the inner crust matter was also considered 
in relation to the cooling of isolated neutron stars~\cite{lattimer,brown} 
and, more recently, in the thermal after-burst relaxation of neutron stars 
from X-ray transients~\cite{Shternin2007,BrownCumming2009,haensel}.

The superfluid properties of the inner crust are essentially determined by
the non-localized neutrons. For baryonic densities smaller than about
1.4 x 10$^{-14}$ g cm$^{-3}$ the non-localized neutrons are
supposed to coexist with nuclei-type clusters~\cite{bbp,negele}. 
At higher densities, before the nuclear matter becomes uniform,
the neutrons and the protons can form other configurations such 
as rods, plates, tubes and bubbles ~\cite{pethick}.

 A microscopic {\it ab initio} calculation of pairing in
 inner crust matter should take into account the polarization effects
 induced by the nuclear  medium upon the bare nucleon-nucleon
 interaction. This is a very difficult task  which is not yet 
 completely solved even for the infinite neutron matter. Thus,
 compared to BCS calculations with bare nucleon-nucleon forces, 
 most of variational or diagrammatic models predict for infinite matter
 a substantial reduction of the pairing correlations due to the 
 in-medium polarisation effects \cite{lombardo}. 
 On the other hand, calculations  based on Monte Carlo techniques
 predict for dilute neutron matter results closer to the 
 BCS calculations (for a recent study see \cite{gezerlis}). 

 A consistent treatment of polarization effects on pairing
 is still missing for inner crust matter (for a recent 
 exploratory study see \cite{baroni}). Therefore at present 
 the most advanced microscopic model applied to inner crust 
 matter remains the Hartree-Fock-Bogoliubov (HFB) approach.
 Pairing correlations have been also considered in the Quasiparticle
 Random Phase Approximation (QRPA)(see Section 2.1 below) in relation
 to the collective modes in inner crust matter \cite{khan}. However,
 a systematic investigation of the effect of collective QRPA excitations
 on thermodynamic properties of inner crust matter is still missing.
 
 The scope of this chapter is to show how the HFB approach
 can be used to investigate the effects of pairing correlations
 on inner crust matter properties. Hence, in the first part
 of the chapter we will discuss the influence of pairing, treated
 in HFB approach at zero temperature, on the structure of inner
 crust matter. Then, using the HFB approach at finite temperature,
 we will show how the pairing correlations affect the specific heat
 and the thermalization of the inner crust matter in the case
 of a rapid cooling scenario.

 In the present study we will focus only to the region of the inner
 crust which is supposed to be formed by a bbc crystal lattice 
 of nuclear clusters embedded in non-localized neutrons. 
The crystal lattice is divided in elementary cells which are treated
in the Wigner-Seitz approximation.

\section{Treatment of pairing in the inner crust of neutron stars}

\subsection{Finite-temperature Hartree-Fock-Bogoliubov approach}
 
In this section we discuss the finite-temperature HFB approximation for a Wigner-Seitz
cell  which contains in its center a nuclear cluster surrounded by a neutron gas. 
The cell contains also relativistic electrons which are considered uniformly distributed.
     
In principle, the HFB equations should be solved by respecting the 
bbc symmetry of the inner crust lattice. However, imposing the exact lattice 
symmetry in microscopic models is a very difficult task (for approximative 
solutions to this problem see Refs.~\cite{chamel2010,Gogelein2007} and the references 
therein). 
We therefore solve the HFB equations for a spherical WS cell, as commonly
done in inner crust studies \cite{negele,baldo2005}. Since we are interested
to describe the thermodynamic properties of the inner crust matter, we 
present here the HFB approach at finite temperature.

The HFB equations for a spherical WS cells have the same form as for
isolated atomic nuclei. Thus, for zero range pairing forces  and 
spherical symmetry, the HFB equations at finite temperature are 
defined as~\cite{goodman1981},
\begin{equation}
\begin{array}{c}
\left( \begin{array}{cc}
h_{T,q}(r) - \lambda_q & \Delta_{T,q} (r) \\
\Delta_{T,q} (r) & -h_{T,q}(r) + \lambda_q 
\end{array} \right)
\left( \begin{array}{c} U_{i,q} (r) \\
 V_{i,q} (r) \end{array} \right) = E_{i,q}
\left( \begin{array}{c} U_{i,q} (r) \\
 V_{i,q} (r) \end{array} \right) ~,
\end{array}
\label{eq:fthfb}
\end{equation}
where $E_{i,q}$ is the quasiparticle energy, $E_{i,q}=\sqrt{(e_{i,q}-\lambda_q)^2+\Delta_{i,q}^2}$, 
$U_{i,q}(r)$ and $V_{i,q}(r)$ are the components of the HFB wave function and $\lambda_q$ is the 
chemical potential ( $q={n,p}$ is the index for neutrons and protons) .
The quantity $h_{T,q}(r)$ is the thermal averaged mean field hamiltonian 
and $\Delta_{T,q}(r)$ is the thermal averaged pairing field.

In a self-consistent HFB calculation based on a Skyrme-type force, as used in the present study,
$h_{T,q}(r)$ and $\Delta_{T,q}(r)$ are expressed in terms of thermal averaged densities, i.e., particle 
density $\rho_{T,q}(r)$, kinetic energy density $\tau_{T,q}(r)$, spin density $J_{T,q}(r)$~
and, respectively, pairing density $\kappa_{T,q}(r)$. 
The thermal averaged densities mentioned above are given by~\cite{sandulescu2004b}:
\begin{eqnarray}
\rho_{T,q}(r) &=&\frac{1}{4\pi} \sum_{i} g_{i,q} \left[  \vert V_{i,q} (r)\vert^2 (1 - f_{i,q} ) + \vert U_{i,q} (r)\vert^2 
f_{i,q} \right]  ,
\label{eq:rhoT}\\
\tau_{T,q}(r) & = & \frac{1}{4\pi} \sum_{i} g_{i,q} \left\{ \left[\left(\frac{dV_{i,q}(r)}{dr}-\frac{V_{i,q}(r)}{r}\right)^2 +
\frac{l_{i,q}(l_{i,q}+1)}{r^2} V_{i,q}(r)^2 \right] (1 - f_{i,q})\right. \nonumber \\
&&\hspace{1.5cm}+ \left. \left[\left(\frac{dU_{i,q}(r)}{dr}-\frac{U_{i,q}(r)}{r}\right)^2 +\frac{l_{i,q}(l_{i,q}+1)}{r^2}
U_{i,q}(r)^2\right]f_{i,q} \right\} , \label{eq:tauT}\\
J_{T,q}(r) & = & \frac{1}{4\pi} \sum_i g_{i,q} \left( j_{i,q}(j_{i,q}+1)-l_{i,q}(l_{i,q}+1)-\frac{3}{4} \right) \nonumber\\
&&\hspace{1.5cm}\times \left( \vert V_{i,q}(r)\vert^2 (1-f_{i,q}) + \vert U_{i,q}(r)\vert^2 f_{i,q} \right) ,
\label{eq:JT}\\
\kappa_{T,q}(r) &=& \frac{1}{4\pi} \sum_{i} g_{i,q} \;U_{i,q}^* (r) V_{i,q} (r) (1 - 2f_{i,q} ) ,
\label{eq:tensor}
\end{eqnarray}
where $f_{i,q} = [1 + \exp ( E_{i,q}/T)]^{-1}$ is the Fermi-Dirac distribution of quasiparticles, 
T is the temperature expressed in energy units, 
and $g_{i,q}=2j_{i,q}+1$ is the degeneracy of the state $i$ with angular momentum $j_i$.
The summations in the equations above are over the spectrum of bound nucleons which form the nuclear cluster
and of unbound neutrons which form the neutron gas. 
A constant density at the edge of the WS cell is obtained
imposing Dirichlet-Von Neumann boundary conditions at the edge of the 
cell~\cite{negele}, i.e., all wave functions of even parity vanish and the derivatives of odd-parity wave functions vanish.

The nuclear mean field has the same expression in terms of densities as in finite
nuclei \cite{dobaczewski}. However, for a WS cell the Coulomb mean field of protons has an additional 
contribution coming from the interaction of the protons with the electrons given by
\begin{equation}
u_\mathrm{Coul}^{pe}(r)=-e^2\int \! d^3r' \, \rho_e(r')\frac{1}{|r-r'|}.
\end{equation}
Assuming that the electrons are uniformly distributed inside the cell, with the
density  $\rho_e=3Z/(4\pi R_{WS}^3)$, one gets
\begin{equation}
u_\mathrm{Coul}^{pe}(r)
=-2\pi e^2\rho_e\left(R_{WS}^2-\frac{1}{3}r^2\right) =
\frac{Ze^2}{2R_{WS}}\left(
\left(\frac{r}{R_{WS}}\right)^2-3\right)
\end{equation}
It can be seen that inside the WS cell the contribution of the
proton-electron interaction to the proton mean field is quadratic in the
radial coordinate.

The pairing field is calculated with a zero range force of the following form
\begin{equation}
V_{{\rm Pair},q}(\mathbf{r_i}, \mathbf{r_j})= V_0 \; g_{{\rm Pair},q}[\rho_{T,n}(\mathbf{r}),\rho_{T,p}(\mathbf{r})] 
(1-P_\sigma) \delta(\mathbf{r}_{ij})\, ,
\label{eq:vpair}
\end{equation}
where $P_\sigma =(1+\hat{\sigma}_1\cdot\hat{\sigma}_2)/2$ is the spin exchange operator.
For this interaction the pairing field is given by
\begin{equation}
\Delta_{T,q}(r)= V_0 \; g_{{\rm Pair},q}[\rho_{T,n}(\mathbf{r}),\rho_{T,p}(\mathbf{r})] \; \kappa_{T,q}(r) .
\label{eq:deltaT}
\end{equation}

In the calculations presented here we use two different functionals for 
$g_{{\rm Pair},q}[\rho_{T,n}(\mathbf{r}),\rho_{T,p}(\mathbf{r})]$.
The first one, called below isoscalar (IS) pairing force, depends  only on the total
baryonic density, $\rho_{T,B}(r)=\rho_{T,n}(r)+\rho_{T,p}(r)$. Its expression is given by
\begin{equation}
g_{{\rm Pair},q}[\rho_{T,n}(\mathbf{r}),\rho_{T,p}(\mathbf{r})]= 1-\eta \left(\frac{\rho_{T,B}(r)}{\rho_0}\right)^\alpha \; ,
\label{eq:is}
\end{equation}
where $\rho_0$ is the saturation density of the nuclear matter.
This effective pairing interaction is extensively used in nuclear structure calculations and it was also 
employed for describing pairing correlations in the inner crust of neutron 
stars~\cite{sandulescu2004a,sandulescu2004b,sandulescu2008,monrozeau}.
The parameters are chosen to reproduce in infinite neutron matter two pairing scenarii, i.e., 
corresponding to a maximum gap of about 3 MeV (strong pairing scenario, hereafter named ISS) and, 
respectively, to a maximum gap around 1 MeV (weak pairing scenario, called below ISW). 
These two pairing scenarii are simulated by two values of the pairing strength, i.e., 
V$_0$=\{-570,-430\} MeV fm$^{-3}$.
The other parameters are taken the same for the strong and the weak pairing, i.e., $\alpha$=0.45, 
$\eta$=0.7 and $\rho_0$=0.16 fm$^{-3}$.
The energy cut-off, necessary to cure the divergence associated to the zero range of the pairing force, 
is introduced through the factor $e^{- E_i/100}$ acting for $E_i > 20$ MeV, where $E_i$ are the 
HFB quasiparticle energies.

\begin{figure}[t]
\begin{center}
\includegraphics[scale=0.7]{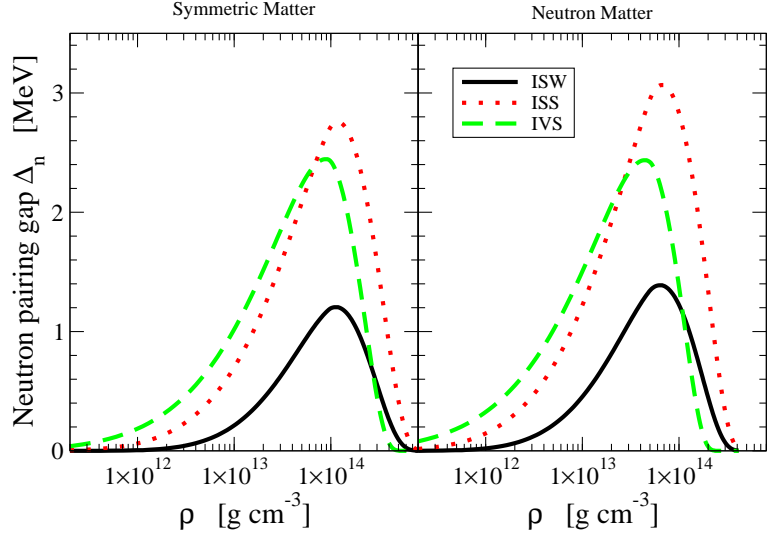}
\caption{(color online) Neutron pairing gap for the interactions
ISW (isoscalar weak), ISS (isoscalar strong)
and IVS (isovector strong)
in symmetric nuclear matter and in neutron matter.}
\label{fig:gap}
\end{center}
\end{figure}

The second pairing functional, referred below as isovector strong pairing (IVS), depends
explicitly on neutron and proton densities and has the following form in the neutron 
channel~\cite{margueron2007b},
\begin{equation}
g_{{\rm Pair},n}[\rho_{T,n}(\mathbf{r}),\rho_{T,p}(\mathbf{r})]= 1 - \eta_s (1-I(r)) \left(\frac{\rho_{T,B}(r)}{\rho_0}\right)^{\alpha_s}
-\eta_n I(r) \left(\frac{\rho_{T,B}(r)}{\rho_0}\right)^{\alpha_n} \; ,
\label{eq:is+iv}
\end{equation}
where $I(r)=\rho_{T,n}(r)-\rho_{T,p}(r)$. 
This interaction is adjusted to reproduce the neutron $^1$S$_0$ pairing gap in neutron and symmetric 
nuclear matter provided by the BCS calculations with the bare nucleon-nucleon forces ~\cite{cao2006}.
In addition, the pairing strength $V_0$ and the cut-off energy are related to each other through the 
neutron-neutron scattering length according to the procedure described in Ref.~\cite{bertsch}.
Therefore this interaction is expected to describe properly the pairing for all the nuclear densities of the 
inner crust matter, including the low density neutron gas.
As shown in Refs.~\cite{margueron2008,bertulani}, this pairing functional describes well the two-neutron
separation energies and the odd-even mass differences in nuclei with open shells in neutrons.
In the present calculations for this pairing functional we have used the parameters  
V$_0$= -703.86~MeV~fm$^{-3}$,  $\eta_s$=0.7115, $\alpha_s$=0.3865, $\eta_n$=0.9727, 
$\alpha_n$=0.3906. The cut-off prescription is the same as for the isoscalar pairing
force.

The pairing gaps in symmetric matter and neutron matter predicted by the three pairing forces
introduced above are represented in Fig.~\ref{fig:gap} for a wide range of sub-nuclear 
densities. It can be seen that the isovector IVS interaction gives a maximum gap closer to the strong
isoscalar ISS force, 
and the ISW interaction predict a suppression of the pairing gap up to saturation density.

To illustrate how the pairing correlations are spatially distributed in the Wigner-Seitz cells
and how they are affected by the temperature, in Fig. 2 are shown the pairing fields for neutrons
in the cells 2 and 5 (see Table 1). It can be noticed that the clusters have a non-trivial influence on
the pairing of neutron gas. Thus, depending of the relative intensity of pairing in the cluster
and the gas region, the presence of the cluster can suppress or enhance the pairing in the
surface region of the cluster.

\begin{figure}[t]
\begin{center}
\includegraphics[scale=0.27,angle=-90.]{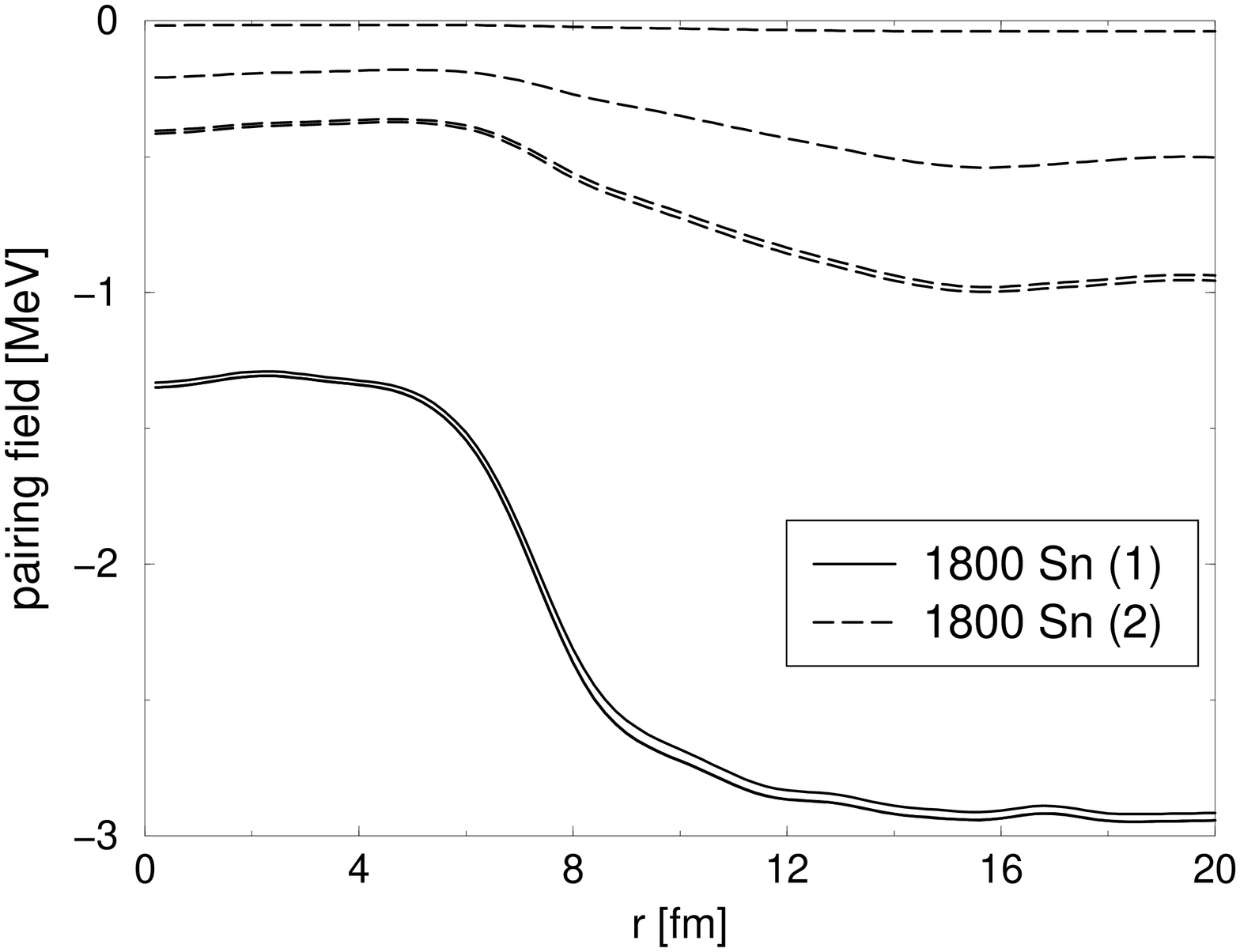}
\includegraphics[scale=0.27,angle=-90.]{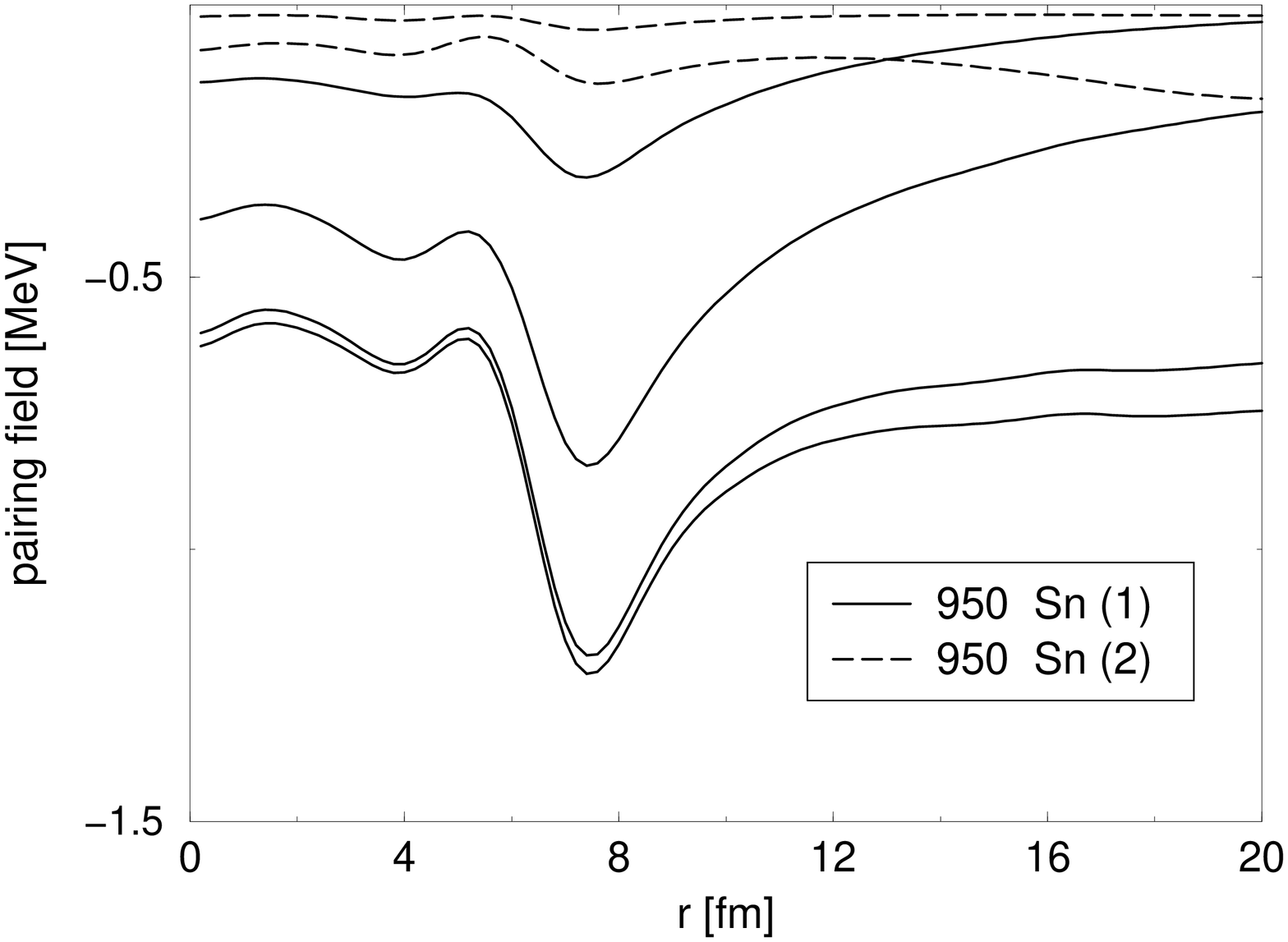}
\caption{Neutron pairing fields for the Wigner-Seitz cells 2 and 5
 (see Table 1) labeled, respectively, as $^{1800}$Sn and $^{950}$Sn.
 The full  and the long-dashed lines correspond to the ISS and ISW pairing
interaction. Here the calculations have been done with the strenghts
$V_0=\{-430,-330\}$ MeV fm$^3$ and with an energy cut-off of 60 MeV.}
\end{center}
\end{figure}

\subsection{Quasiparticle Random Phase Approximation (QRPA)} 

 Pairing correlations  affect not only the ground state properties
 of inner crust matter but also its excitations modes. The non-collective
 excitations are commonly described by the quasiparticle energies
 obtained solving the HFB equations. To calculate
 the collective excitations one needs to take into account
 the residual interaction between the quasiparticles. In what follows
 we discuss briefly the collective modes of the inner crust
 matter in the framework of QRPA, which takes properly into account
 the pairing correlations \cite{khan}.
  
 The QRPA can be obtained from the time-dependent HFB approach in the
 limit of linear response. In the linear response theory the fundamental
 quantity is the Green function which satisfies the Bethe-Salpeter equation 
\begin{equation}\label{eq:bs}
\bf{G}=\left(1-\bf{G}_0\bf{V}\right)^{-1}\bf{G}_0=\bf{G}_0+\bf{G}_0\bf{V}\bf{G}.
\end{equation}
The unperturbed Green's function $\bf{G}_0$ has the form:
\begin{equation}\label{eq:g0}
{\bf{G}_0}^{\alpha\beta}({\bf r}\sigma,{\bf r}'{\sigma}';\omega)=
\sum_{ij} \hspace{-0.5cm}\int \hspace{0.3cm} \frac{{\cal U}^{\alpha 1}_{ij}({\bf r}\sigma)
\bar{{\cal U}}^{*\beta 1}_{ij}({\bf r}'\sigma')}{\hbar\omega-(E_i+E_j)+i\eta}
-\frac{{\cal U}^{\alpha 2}_{ij}({\bf r}\sigma)
\bar{{\cal U}}^{*\beta 2}_{ij}({\bf r}'\sigma')}{\hbar\omega+(E_i+E_j)+i\eta},
\end{equation}
where $E_i$ are the HFB quasiparticle energies and ${\cal U}_{ij}$ are 3 by 2
matrices expressed in term of the two components of the HFB wave functions \cite{khan_qrpa}. 
The $\sum$ \hspace{-0.45cm}$\int$ ~ symbol in the equation above indicates that the
summation is taken over the bound and unbound quasiparticle states. The latter corresponds
here to the non-localised neutrons in the WS cell.

\begin{figure}[t]
\begin{center}
\hspace{1cm}
\includegraphics[scale=0.5]{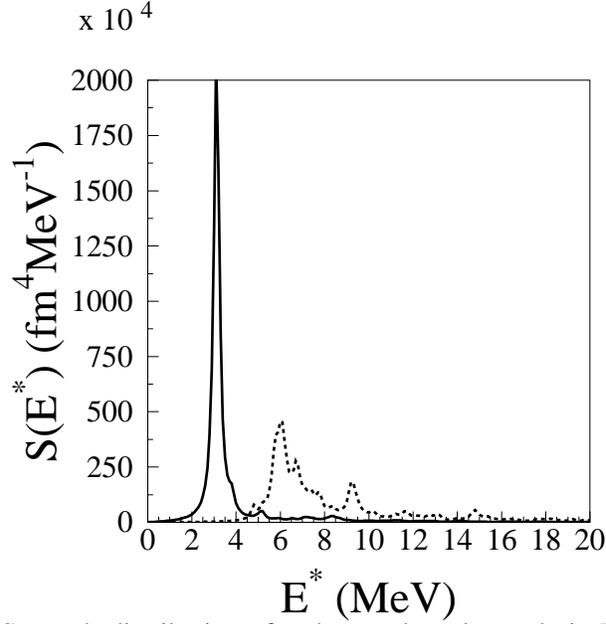}
\caption{Strength distributions for the quadrupole mode in WS cell 2 of Table 1. The full (dashed)
line corresponds to the QRPA (HFB) strength. The results are for ISS pairing force with the same 
parameters as used in Figure 1.}
\end{center}
\end{figure}

$\bf{V}$ is the matrix of the residual interaction 
expressed in terms of the second derivatives of the HFB energy 
functional, namely:					   
\begin{equation}\label{eq:vres}
{\bf{V}}^{\alpha\beta}({\bf r}\sigma,{\bf r}'{\sigma}')=
\frac{\partial^2{\cal E}}{\partial{\bf{\rho}}_\beta({\bf r}'{\sigma}')
\partial{\bf{\rho}}_{\bar{\alpha}}({\bf r}\sigma)},~~~\alpha,\beta = 1,2,3.
\end{equation}
In the above equation $\{\rho_1,\rho_2,\rho_3\} \equiv \{\rho, \kappa, \kappa^* \}$,
where $\rho$ and $\kappa$ are, respectively, the particle~(2) and pairing~(5) densities;
the notation $\bar{\alpha}$ means that whenever $\alpha$ is 2 or 3 
then $\bar{\alpha}$ is 3 or 2.

The linear response of the system to external perturbation is commonly
described by the strength function. Thus, when the external perturbation
is induced by a particle-hole external field $F$ the strength function writes:
\begin{equation}\label{eq:stren}
S(\omega)=-\frac{1}{\pi}Im \int  F^*({\bf r}){\bf{
G}}^{11}({\bf r},{\bf r}';\omega) F({\bf r}')
d{\bf r}~d{\bf r}'
\end{equation}
where ${\bf{G}}^{11}$ is the (ph,ph) component of the QRPA Green's function.

As an example in Fig. 3 it is shown the strength function for the quadrupole response calculated for the
WS cell 2 of Table~1 below~\cite{khan}.
The results correspond to the isoscalar pairing force with the strength $V_0$=-430 MeV fm$^{-3}$ 
and an energy cut-off of 60 MeV.
It can be seen that the unperturbed spectrum, 
distributed over a large energy region, becomes concentrated almost entirely in the peak 
located at about 3 MeV when the residual interaction between the quasiparticles is introduced. 
The peak collects more than 99$\%$ of the total quadrupole strength and it is extremely collective.
An indication of the extreme collectivity of this low-energy mode can be
also seen from its reduced transition probability, B(E2), which is equal
to $25\times10^3$ Weisskopf units. 
This value of B(E2) is two orders of magnitude higher than in standard nuclei.
This underlines the fact that in this WS cell the collective dynamics of the neutron gas dominates
over the cluster contribution.
In Ref.~\cite{khan} it is shown that similar collective modes appears for the monopole and the dipole excitations. 
A very collective low-energy quadrupole mode it was also found in all the Wigner-Seitz cells with 
Z=50~\cite{Grasso2008}. However, a systematic investigation of the influence of these collective modes on 
the thermodynamic properties of inner crust matter is still missing.

\section{The effect of pairing on inner crust structure}
\label{sec:structure}

The first microscopic calculation of the inner-crust structure
was performed by Negele and Vautherin in 1973~\cite{negele}.
In this work the crystal lattice is divided in spherical cells which are treated in the
Wigner-Seitz (WS) approximation. The nuclear matter from each cell is described in the
framework of Hartree-Fock (HF) and the pairing is neglected.
The properties of the WS cells found in Ref.~\cite{negele}, determined for a
limited set of densities, are shown in Table~I.
The most remarkable result of this calculation is that the majority of the cells
have semi-magic and magic proton numbers, i.e., Z=40,50.
This indicates that in these calculations there are strong proton shell effects,
as in isolated atomic nuclei.

\begin{table}[t]
\setlength{\tabcolsep}{.13in}
\renewcommand{\arraystretch}{0.9}
\begin{tabular}{cccccc}
\hline
   $N_{cell}$ & $\rho_B$ & $k_F$  & $N$ & $Z$ & $R_{WS}$ \\
    & [g cm$^{-3}$] & [fm$^{-1}$] & & & [fm] \\
\hline\hline
  1 & $7.9\ 10^{13}$  & 1.12 &1460 &40  &19.7   \\
  2 & $3.4\ 10^{13}$  & 0.84 &1750 &50  &27.7  \\
  3 & $1.5\ 10^{13}$  & 0.64 &1300 &50  &33.2  \\
  4 & $9.6\ 10^{12}$  & 0.55 &1050 &50  &35.8   \\
  5 & $6.2\ 10^{12}$  & 0.48 &900 &50  & 39.4  \\
  6 & $2.6\ 10^{12}$  & 0.36 &460 &40  &42.3   \\
  7 & $1.5\ 10^{12}$  & 0.30 &280 &40  &44.4   \\
  8 & $1.0\ 10^{12}$  & 0.26 &210 &40  &46.5   \\
  9 & $6.6\ 10^{11}$  & 0.23 &160 &40  &49.4 \\
 10 & $4.6\ 10^{11}$ & 0.20 &140 &40  &53.8  \\
\hline
\end{tabular}
\caption{The structure of the Wigner-Seitz cells obtained from a density matrix expansion (DME)~\cite{negele}
$\rho_B$ is the baryon density, $k_F=(3\pi^2 n_B)^{1/3}$ the Fermi momentum calculated as in 
Ref.~\cite{baldo2005} where $n_B$ is the number of baryons per fm$^3$, N and Z are the numbers of 
neutrons and protons while $R_{WS}$ is the radius of the WS cells.
Compared to Ref. \cite{negele} here it is not shown the cell with the
highest density located at the interface with the pasta phase. }
\label{tab:NV}
\end{table}

The effect of pairing correlations on the structure of Wigner-Seitz cells was first investigated
in Refs.~\cite{baldo2005,baldo2006,baldo2007} within the Hartree-Fock BCS (HFBCS) approach.
In this section we shall discuss the results of a recent calculations based 
on  Hartree-Fock-Bogoliubov approach \cite{grill2011}. This approach offers better
grounds than HF+BCS approximation for treating pairing correlations
in non-uniform nuclear matter with both bound and unbound neutrons.

As in Ref.~\cite{negele}, the lattice structure of the inner crust is described as a set
of independent cells of spherical symmetry treated in the WS approximation.
For baryonic densities below $\rho \approx 1.4 \times 10^{14}$ g/cm$^3$, 
each cell has in its center a nuclear cluster (bound protons and neutrons)
surrounded by low-density and delocalized neutrons and immersed in a uniform
 gas of ultra-relativistic electrons which assure the charge neutrality. 
At a given baryonic density the structure of the cell,
i.e., the N/Z ratio and the cell radius is determined from the minimization
over N and Z of the total energy under the condition of beta equilibrium.
The energy of the cell, relevant for determining the cell structure, has contributions from
the nuclear and the Coulomb interactions. Its expression is written in the following form
\begin{equation}
E = E_M + E_N+T_e+E_L .
\label{eq:energy}
\end{equation}
The first term is the mass difference $E_M=Z(m_p+m_e)+(N-A)m_n$
where N and Z are the number of neutrons and protons in the cell
and A=N+Z. $E_N$ is the binding energy of the nucleons, which
includes the contribution of proton-proton Coulomb interaction
inside the nuclear cluster. $T_e$ is the kinetic energy of the
electrons while $E_L$ is the lattice energy which takes into
account the electron-electron and electron-proton interactions.
The contribution to the total energy coming from the interaction
between the WS cells \cite{oyamatsu} it is not considered since
it is very small compared to the other terms of Eq.(1).

We shall now discuss the effect of pairing correlations on the
structure of the WS cells. To study the influence of pairing
correlations we have performed HFB calculations with the three
pairing interactions introduced in Section 2.1.

\begin{figure}[ht]
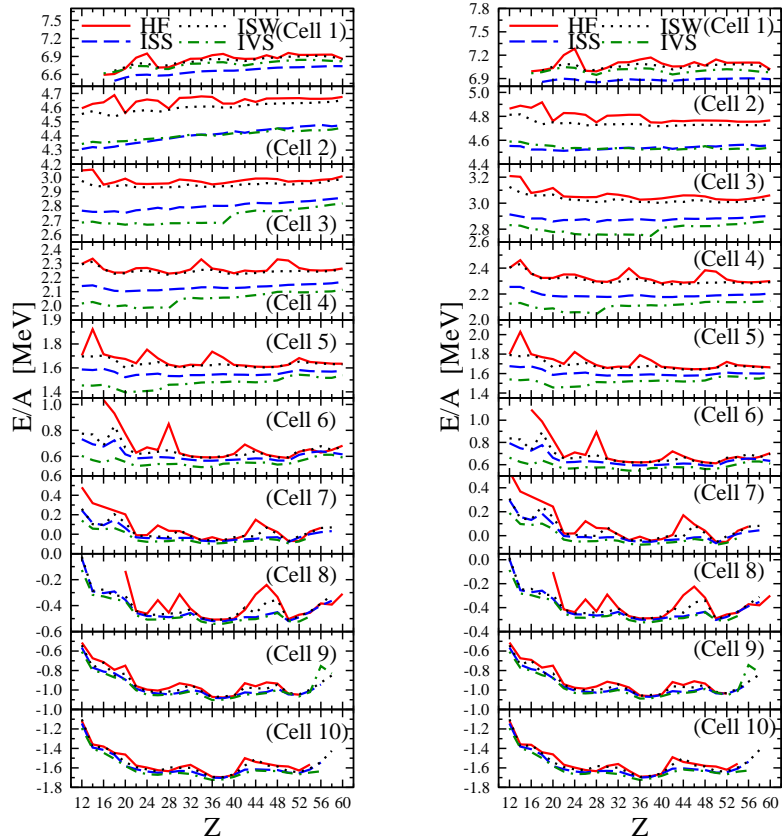

\begin{center}
\includegraphics[scale=0.6]{PRCXX-fig_ehfb.eps}
\hspace{1.0 cm}
\includegraphics[scale=0.6]{PRCXX-fig_ehfbc.eps}
\caption{(color online) The HFB energies per particle as function of
proton number for the pairing forces ISW (dotted line),
ISS (dashed line) and IVS (dashed-dotted line). The solid lines
represent the HF results. In the left pannel are shown the results
obtained including the finite size corrections.}
\label{fig:ehfb}
\end{center}
\end{figure}

The structure of the WS cells obtained in the HFB approach
is given in Table~II while in  Fig.~\ref{fig:ehfb} it is
shown the dependence of the binding energies, at beta equilibrium,
on protons number. The contribution of pairing energy, nuclear energy
and electron kinetic energy to the total energy are given in Fig. 5
for the cells 2 and 6. From this figure we can notice that the nuclear
binding energy is almost compensated by the kinetic energy of the
electrons which explains the weak dependence of the total energy
on Z seen in Fig. 4. In Fig.4 we observe also that pairing is smoothing
significantly the variation of the HF energy with Z. For this reason
the HFB absolute minima are very little pronounced compared to the other
neighboring energy values.

\begin{table}[ht]
\renewcommand{\arraystretch}{0.9}
\begin{tabular}{c|cccc|cccc|cccc}
\hline $N_{cell}$ & \multicolumn{4}{|c|}{$N$} & \multicolumn{4}{|c|}{$Z$} & \multicolumn{4}{|c}{$Z_{corr}$}\\
 & HF & ISW & ISS & IVS & HF &ISW & ISS & IVS & HF &ISW & ISS & IVS \\
\hline \hline
  2 &       &       &       &       &      &      &      &      & $40$ & $40$ & $22$ & $42$ \\
  3 & $318$ & $514$ & 442 & 554 & $16$ & $24$ & 20 & 24 & $54$ & $40$ & $28$ & $38$ \\
  4 & $476$ & $534$ & $382$ & $570$ & $28$ & $28$ & $20$ & $28$ & $40$ & $40$ & $40$ & $28$ \\
  5 & $752$ & $320$ & $328$ & $344$ & $46$ & $20$ & $20$ & $20$ & $46$ & $48$ & $44$ & $20$ \\
  6 & $454$ & $428$ & $374$ & $346$ & $50$ & $48$ & $36$ & $34$ & $50$ & $50$ & $50$ & $34$ \\
  7 & $316$ & $344$ & $324$ & $240$ & $50$ & $50$ & $50$ & $36$ & $50$ & $50$ & $50$ & $50$ \\
  8 & $174$ & $220$ & $186$ & $174$ & $36$ & $50$ & $38$ & $36$ & $36$ & $50$ & $38$ & $36$ \\
  9 & $120$ & $112$ & $128$ & $116$ & $38$ & $36$ & $38$ & $36$ & $38$ & $36$ & $38$ & $36$ \\
 10 & $ 94$ & $ 82$ & $ 90$ & $ 82$ & $38$ & $36$ & $38$ & $36$ & $36$ & $36$ & $36$ & $36$ \\
\hline
\end{tabular}
\caption{The structure of Wigner-Seitz cells obtained in the HF and HFB
approximation. The results corresponds to the isoscalar weak
(ISW), isoscalar strong (ISS) and isovector-isoscalar (IVS)
pairing forces. In the last 4 columns are shown the proton numbers  obtained
with the finite size corrections. In the table are shown only the structures
of the cells which could be well-defined by the present calculations.} 
\end{table}

From Fig.~\ref{fig:ehfb} we observe that in the cells 1-2 the binding energy
does not converge to a minimum at low values of Z. For the cells 3-4, 
although absolute minima can be found for HF or/and HFB calculations,
these minima are very close to the value of binding energy at the 
lowest values of Z we could explore. Therefore the structure of these 
cells is ambiguous. The situation is different in the cells 5-10 where
the binding energies converge to absolute minima which are well below the
energies of the configurations with the lowest Z values.

\begin{figure}[ht]
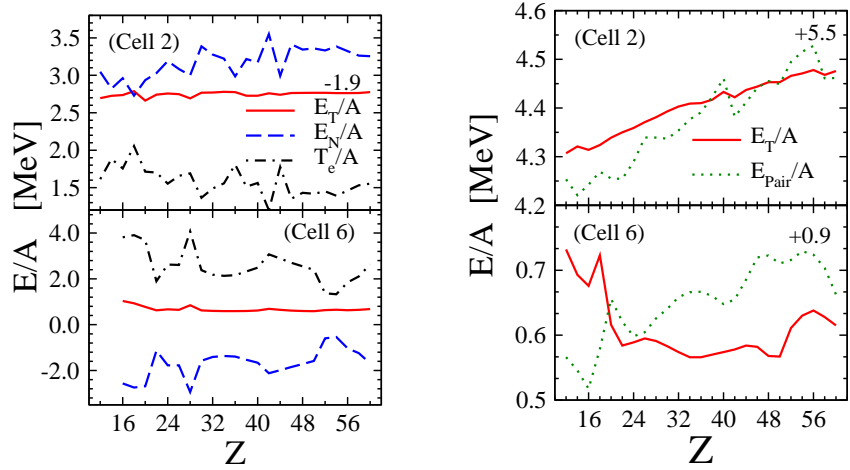

\begin{center}
\includegraphics[scale=0.6]{PRCXX-fig_ehf1.eps}
\hspace{1.0 cm}
\includegraphics[scale=0.6]{PRCXX-fig_ehf2.eps}
\caption{(color online) The different contributions to the total
energy in the cells 2 and 6 for the HF (left pannel) and HFB
(right pannel) calculations. Are shown: the total energy (solid line),
the nuclear energy (dashed line), the kinetic energy of the electrons
(dashed-dotted line), and the pairing energy for the ISS pairing
interaction (dotted line). The pairing energies are shifted up
as indicated in the figure.}
\label{fig:ehf2}
\end{center}
\end{figure}

From Table II it can be observed that the structure of some cells becomes
very different when the pairing is included. Moreover, these differences
depend significantly on the intensity of pairing (see the 
results for ISW and ISS). However, as seen in Fig. 4, for the majority
 of cells the 
absolute minima are very little pronounced relative to the other energy
values, especially for the HFB calculations. Therefore one cannot  draw
clear quantitative conclusions on how much pairing is changing the 
proton fraction in the cells.

When the radius of the cell becomes too small the boundary conditions
imposed at the cell border through the WS approximation generate an
artificial large distance between the energy levels of the nonlocalized
neutrons. Consequently, the binding energy of the neutron gas is significantly
underestimated. An estimation  of how large could be the errors in
the binding energy induced by the WS approximation can be obtained
from the quantity
\begin{equation}
f(\rho_{n},R_{WS}) \equiv
B_{inf.}(\rho_{n}) - B_{WS-inf.}(\rho_{n},R_{WS}) \; ,
\end{equation}
where the first term is the binding energy per neutron for infinite neutron
matter of density $\rho_n$ and the second term is the binding energy of neutron
matter with the same density calculated inside the cell of radius $R_{WS}$
and employing  the same boundary conditions as in HF or HFB calculations.
In Ref.~\cite{margueron2007} it was proposed for the finte size energy 
correction, Eq. 26, the following parametrisation
\begin{equation}
f(\rho_{n_g},R_{WS}) = 89.05 (\rho_{n_g}/\rho_0)^{0.1425}
R_{WS}^{-2} \; ,
\end{equation}
where $\rho_{n_g}$ is the average density of neutrons in the gas
region extracted from a calculation in which the cell contains
both the nuclear cluster and the nonlocalized neutrons while
$\rho_0$ is the nuclear matter saturation density.

How the energy corrections  described by Eq. (18) influences the HF (HFB)
results can be seen in Fig. 4 (right pannel) and in Table II (last four columns).
 As expected, the influence of the corrections is more important for
the cells 1-5, in which the neutron gas has a higher density, and
for those configurations corresponding to small cell radii

\section{Effect of pairing on the thermodynamic properties in the inner crust matter}

\subsection{Specific heat of the baryonic matter}
\label{sec:specificheat}

\begin{figure}[t]
\begin{center}
\includegraphics[scale=1.3,angle=0]{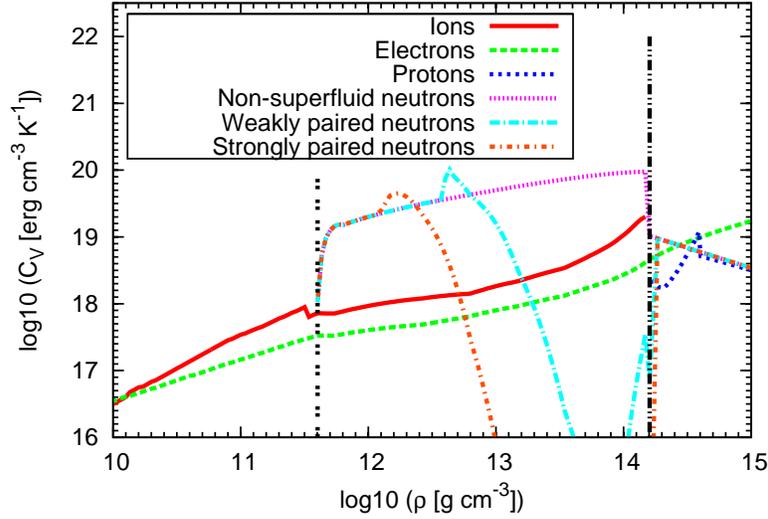}
\caption{Specific heat in the crust of the neutron star for the different components 
of the star matter and for a temperature of 10$^9$~K.}
\label{fig:cv}
\end{center}
\end{figure}

The specific heat of the inner crust has contributions from the electrons, the lattice (or Ion)
and the baryonic matter (essentially the de-localized neutrons). 
The contribution of these 3 components is shown in Fig.~\ref{fig:cv} where different
scenarios for the neutrons are considered.
If neutrons are non-superfluid they dominate the contribution of the other component in the
inner crust. The effect of the neutron superfluidity is the reduce strongly the contribution of the
neutrons.
It is therefore expected a very large effect induced by the neutron superfluidity that we now discuss.
The specific heat of the neutrons 
is calculated from the quasiparticle energies obtained solving  the HFB  equations
at finite-temperature in the Wigner-Seitz approximation, as discussed in Section 2.1.
For the WS cells we use the structure determined in Ref.\cite{negele} (see Table 1).

The specific heat of the neutrons is calculated by 
\begin{equation}
C_{V}=\frac{T}{V}\frac{\partial S}{\partial T} , 
\end{equation}
where V is the volume of the Wigner-Seitz cell, T is the temperature and S is
the entropy. The latter is given by
\begin{equation}
S=-k_B \sum_{i,q} g_{i,q} (f_{i,q} \ln f_{i,q}+(1-f_{i,q})\ln (1-f_{i,q})).
\end{equation}

\begin{figure}[t]
\begin{center}
\includegraphics[scale=0.2,angle=-90]{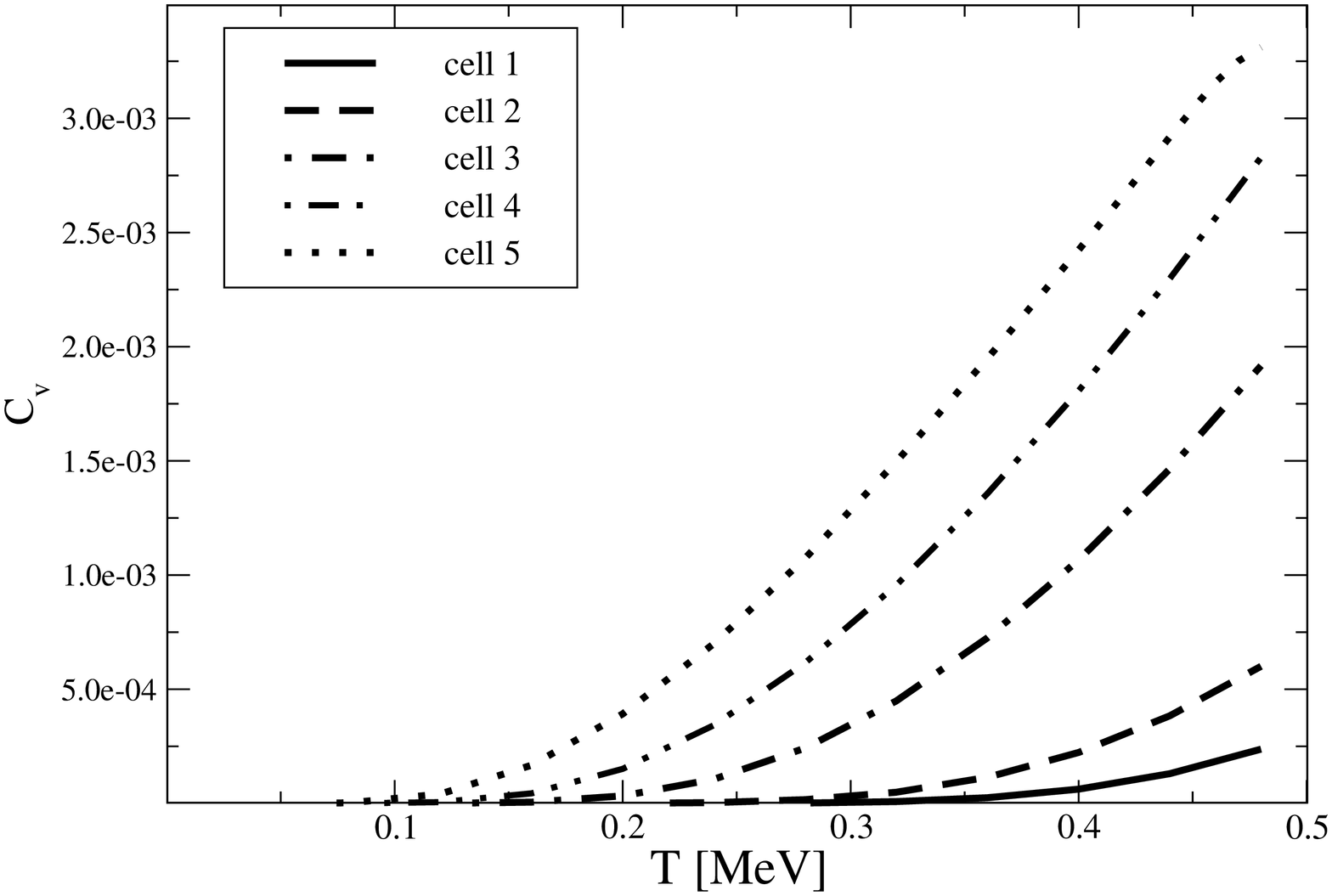}
\includegraphics[scale=0.2,angle=-90]{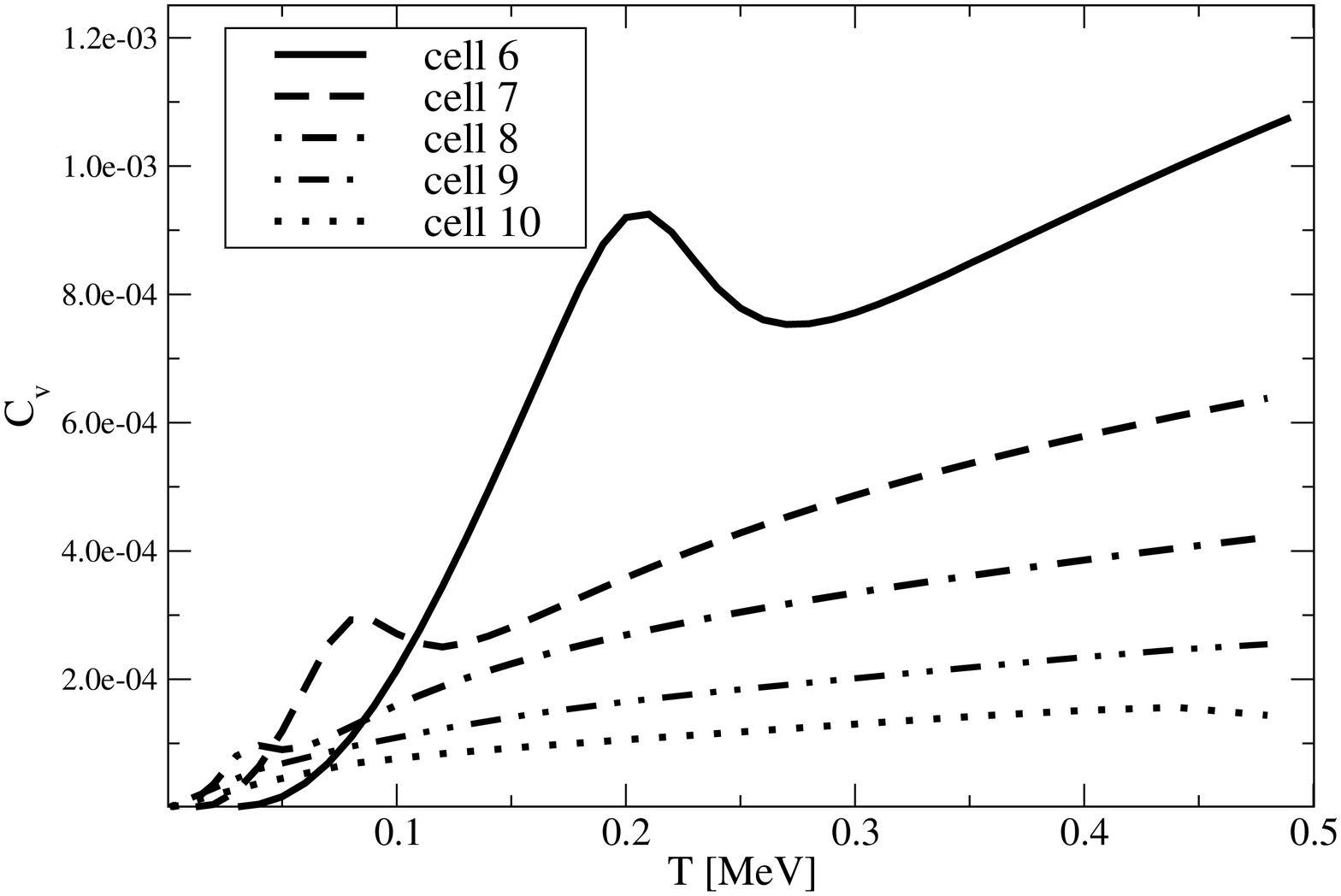}
\caption{Specific heat of neutrons in various WS cells for isoscalar strong ISS pairing.
The specific heat is given in units of Boltzmann constant $k_B$.}
\end{center}
\end{figure}

\begin{figure}[t]
\begin{center}
\includegraphics[scale=0.35,angle=0]{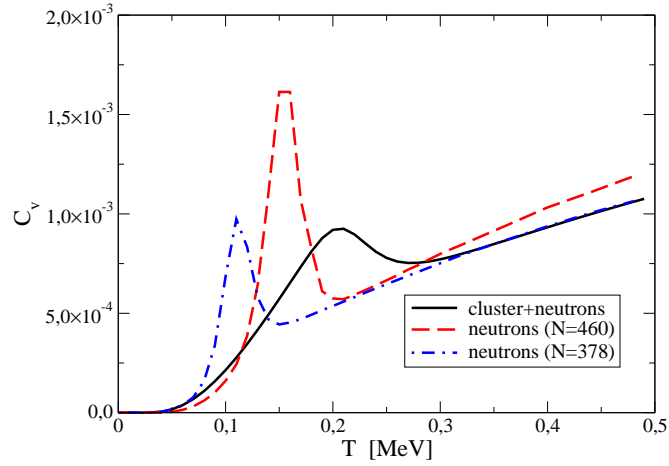}
\caption{Neutron-specific heat in the the WS cell 6 for isoscalar strong pairing. 
The results corresponds to various approximations discussed in the paper. 
The specific heat is given in units of Boltzman constant $k_B$.}
\label{fig:fig3}
\end{center}
\end{figure}

In Fig. 7 are shown the results obtained for the specific heat in the case of a strong
isoscalar pairing interaction.
To illustrate the particular behavior of the specific heat in non-uniform matter and the validity of 
various approximations, in what follows we shall discuss in more detail the results for the cell 
number 6, which contains N=460 neutrons and Z=40 protons (see Table~1). 
In this cell the HFB calculations predict 378 unbound neutrons. 
The specific heat given by the HFB spectrum, in which the contribution of the cluster is included, 
is shown in Fig.~\ref{fig:fig3} by full line. 
In the same figure are shown also the specific heats corresponding to two approximations
employed in some studies~\cite{lattimer,pizzochero,page}. 
In these approximations the non-uniform distribution of neutrons is replaced with a uniform 
gas formed either by the total number of neutrons in the cell (N=460, dashed line), 
or by taking only the number of the unbound neutrons (N=378, dashed-dotted line). 
How these approximations work is seen in Fig.~\ref{fig:fig3}. 
To make the comparison meaningful, the calculations for the uniform neutron gas are 
done solving the HFB equations with the same boundary conditions as for the non-uniform 
system, i.e.,  neutrons+cluster. As seen in Fig.~\ref{fig:fig3}, the transition
from the superfluid phase to the normal phase is taking place at a lower temperature in the
case of uniform neutron gas, especially when are considered only the unbound neutrons. 
The critical temperature is therefore lower in uniform matter (for the two prescriptions often used)
than in non-uniform matter.

\subsection{Pairing and thermalization time of the inner crust}
\label{sec:thermalization}

Several studies have shown that the thermalization time depends  significantly on
the superfluid properties of the inner crust baryonic matter~\cite{lattimer,gnedin2001,pizzochero,
monrozeau,sandulescu2008}. 
This dependence is induced through the specific heat of unbound neutrons, 
strongly affected by the pairing energy gap.
As an example we refer to Fig. 6 where it is seen that neutron superfluidity strongly suppress the 
neutron specific heat which,
in the absence of the pairing would dominate over the specific heat of  electrons and the lattice.

\begin{figure}[t]
\begin{center}
\includegraphics[scale=1.2]{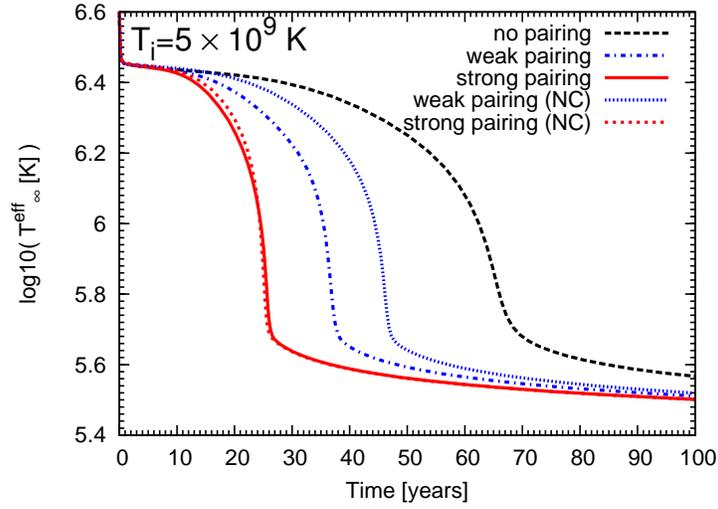}
\caption{Time evolution of the apparent surface temperature for the 
initial temperatures $T_i=500$~keV. NC indicates the results
of the calculations obtained by neglecting  the effect of the clusters.}
\label{fig:tsurf}
\end{center}
\end{figure}

To estimate the effect of neutron superfluidity to the thermalization time of inner
crust matter we use a rapid cooling scenario in which the core arrives 
quickly to a much smaller temperature than the crust due to the direct URCA
colling process.
The thermalization time is defined as the time for the core nd the crust temperatures to
equilibrate.
The heat diffusion is described by the relativistic heat equation~\cite{thorne1977}:
\begin{equation} 
\frac{\partial}{\partial r} \left[ \frac{K r^2}{\Gamma(r)} e^\phi 
\frac{\partial }{\partial r} (e^\phi T) \right] = 
r^2 \Gamma(r) e^\phi \left( C_V \frac{\partial T}{\partial t}
+ e^{\phi}Q_{\nu} \right),
\label{eq:heat}
\end{equation}
where $t$ is the time, $K$ is the thermal conductivity, $C_V$ is the specific heat and $Q_\nu$ is the 
neutrino emissivity. 
The effect of the gravity is given through the  gravitational potential $\phi$, which enters in the
definition of the redshifted temperature $\tilde{T} = Te^{\phi}$, and the quantity
$\Gamma(r)=\left( 1-2Gm(r)/rc^2 \right)^{-1/2}$, where  $G$ is the 
gravitational constant and $m(r)$ is the gravitational mass included 
in a sphere of radius $r$. 
The latter is obtained from the Tolman-Oppenheimer-Volkoff (TOV) equations~\cite{haensel} based on 
an equation of state obtained from SLy4 Skyrme interaction~\cite{douchin}. 
More details on the cooling model are given in Ref. \cite{fortin2010}.

The time evolution of the apparent surface temperature $T^\mathrm{eff}_\infty$ is displayed 
in Fig.~\ref{fig:tsurf} for the initial temperatures of the crust $T_i=500$~keV \cite{fortin2010}. 
The effective surface temperature shown in Fig.~\ref{fig:tsurf} is obtained from the temperature 
at the bottom of the crust, $T_\mathrm{b}=T(\rho_\mathrm{b})$, where 
$\rho_\mathrm{b}= 10^{10}$ g.cm$^{-3}$, using the relationship given in Ref.~\cite{potekhin1997} 
for a non-accreted envelope. The results shown in Fig.~\ref{fig:tsurf} corresponds 
to a neutron star of mass 1.6~$M_\odot$ in which the inner crust extends from $R_c$=10.72~km, 
which is the radius at the core-crust interface, to 11.19~km.
As can be noticed, the pairing enhances significantly the cooling at the surface of the star. 
In Fig.~\ref{fig:tsurf} are also shown the apparent surface temperatures  obtained neglecting the 
effect of the clusters, i.e., supposing that the neutron specific heat is given solely by that
of the neutron gas. In this case the neutron specific heat is calculated from the 
quasiparticle spectrum of BCS equations solved for infinite neutron matter at a density corresponding
to that of the external neutrons in  the WS cell~\cite{levenfish1994}.
In the case of weakly pairing scenario (ISW), the apparent surface temperature is dropping faster for superfluid 
non uniform matter than for superfluid uniform matter. 
For the strong pairing scenario, since the pairing correlations suppress the role of the neutron in 
almost the entire inner crust, the effect of the clusters is less important ~\cite{fortin2010}.

\begin{figure}[t]
\begin{center}
\includegraphics[scale=0.4,angle=-90]{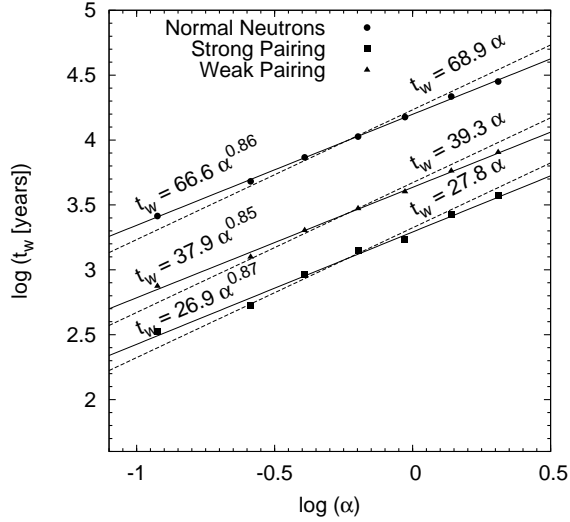}
\caption{Cooling times $t_w$ versus scaling parameter $\alpha$ for three pairing scenarii as discussed
in the text. The results correspond to neutron stars with masses between 1.4 and 2.0 $M_{\odot}$. 
The fitting curves are given for the case of a linear scale (dashed lines, right side) and for a fractional power 
of the scaling parameter $\alpha$ (solid lines, left side).}
\label{fig:ct}
\end{center}
\end{figure}

A simple random walk model for the  cooling is sometimes used to estimate the thermalization 
time~\cite{brown,pizzochero,monrozeau}.
In this model the diffusion of the cooling wave towards the core is calculated without taking into 
account the dynamical change of the temperature through the whole crust and neglecting the neutrino emissivity in Eq.~(\ref{eq:heat}).
The crust is divided into shells of thickness $R_i$ for which the thermal diffusively $D_i=K_i/C_{V,i}$ 
is considered as constant. 
From a dimensional analysis of the heat equation~(\ref{eq:heat}), the relaxation time of each of these 
shells is defined as, 
\begin{equation}
\tau_{i}=\Gamma(R)^3 R_{i}^2/(\gamma D_i),
\label{eq:taui}
\end{equation}
where $\gamma$ is a geometrical factor and $R$ is the radius of the neutron star.
The total relaxation through the crust is given by,
\begin{equation}
\tau_{th}=\left( \sum_i \sqrt{ \tau_i}\right)^2 .
\label{eq:tauth}
\end{equation}
The thermalization time defined as (\ref{eq:tauth}) satisfy the condition that in a uniform density case, 
$\tau_{th}$ is independent of the shell discretization.
Using this model, the effect of the neutron superfluidity has also been estimated to be 
large~\cite{pizzochero,monrozeau}, and is qualitatively similar to the result deduced from solving 
of the heat diffusion equation~(\ref{eq:heat}).

In some previous studies \cite{lattimer,gnedin2001} it was found that the thermalization time,
defined as the time needed for the crust to arrive to the same temperature with the core, 
satisfies the scaling relation $t_{th}= t_1 \alpha^\beta$ where 
\begin{equation}
\alpha=\left(\frac{\Delta R_{\textrm{crust}}}{\textrm{1 km}}\right)^2\left(1-2GM/Rc^2\right)^{-3/2}
\end{equation}
depends solely on the global properties of the neutron star, i.e., 
the crust thickness $\Delta R_{\textrm{crust}}$, 
the star radius $R$ and the mass of the star $M$. 
The scaling relation can be inferred from the random walk model presented above.
In Fig.~\ref{fig:ct} are shown as a function of the scaling parameter $\alpha$ the cooling times 
we have obtained from the solution of the heat diffusion equation~(\ref{eq:heat}) for various
neutron stars masses lying between 1.4 and 2.0 $M_{\odot}$ with a step of $0.1 M_{\odot}$. 
In this figure are displayed two sets of fits for $t_w$ versus $\alpha$, i.e., 
a linear fit with $\beta=1$ (dashed line) and a fit with a fractional value for $\beta$ (solid line).
It can be noticed that the best fit is obtained with a fractional value for $\beta$, which is equal to 0.86 for the 
normal neutrons, to 0.85 for weakly paired neutrons and to 0.89 for strongly paired neutrons.
Considering the simpler linear fit, i.e., $t_w \approx \alpha t_1$, as done in ~\cite{lattimer,gnedin2001},
we get for the normalized time $t_1$ the values $t_1=\lbrace{68.9,39.3,22.3}\rbrace$ corresponding, 
respectively, to the normal neutrons, neutrons with weak pairing and neutrons with strong pairing. 
For a $1.5$~$M_\odot$ neutron star with $\alpha=1.15$, we obtain $t_1=\lbrace{66.4,37.4,26.6}\rbrace$.
These values for $t_1$ are larger than that of Ref.~\cite{gnedin2001} (Table 2) by a factor 2.3 in the 
non-superfluid case, and 3.4 (3.0) for the weak (strong) pairing scenario. 
These differences could be explained by the effects of the nuclear clusters on the neutron specific heat, 
disregarded in Ref.~\cite{gnedin2001}, and by different neutrinos processes and thermal conductivities 
in the core matter used in the two calculations.

\section{Summary and Conclusions}
\label{sec:conclusion}

In the first part of this chapter we have discussed  the influence
of pairing correlations on the structure of inner crust of neutron stars. 
The study was done for the region of the inner crust which is supposed to
be formed by a lattice of spherical clusters embedded in a gas of
neutrons. The lattice was treated as a set of independent cells described 
in the Wigner-Seitz approximation. To determine the structure of a cell
we have used the nuclear binding energy given by the Skyrme-HFB approach.
 For the cells with high density and small radii the binding energies do
not converge to a minimum when the proton number has small values.
We believe that 
it is related to the discretization of the continuum.
For a small radius of the cell
the average distance between the energy levels of the non-localised
neutrons becomes artificially large which cause an underestimation
of the binding energy. To correct this drawback we have used an empirical
expression based on the comparison between the binding energy of neutrons
calculated in infinite matter and in a finite-size spherical cell~\cite{margueron2007}. 
We found that the finite size corrections to the binding energies
are significant for the high density cells with small proton numbers. 
This conclusion indicates the need of a more accurate evaluation of
the errors induced by the finite size of the WS cells on nuclear binding
energy, which requires to go beyond the empirical expression used here.

In the second part we have discussed how thermalization of neutron
stars crust depends on pairing properties and on  cluster structure of
 the inner crust matter. The thermal evolution was obtained by solving the 
relativistic heat equation with initial conditions specific to a rapid cooling process.
The specific heat of neutrons was calculated from the HFB spectrum.
The results show that the crust thermalization is strongly influenced
by the intensity of pairing correlation.
It is also shown that the cluster structure of the inner crust affects significantly the time 
evolution of the surface temperature.

The thermodynamic properties of the inner crust discussed in this chapter
are based on the excitation spectrum of HFB equations. As it was shown in 
Section 2.2, the QRPA approach predicts in some cells  collective excitations
located at low energies, comparable with the pairing energy gap. It is thus
expected a significant contribution of the collective modes to the specific
heat and to the thermalization process of inner crust matter.
This is an open issue which deserves further investigations.

\bigskip

{\bf Acknowledgements:}
We thank to our young collaborators, Fabrizzo Grill and Morgane Fortin for having
done most of the calculations presented in Sections 3 and 4. The work presented
in this review paper was partially financed by the European Science Foundation
through the project "New Physics of Compact Stars", by the Romanian Ministry 
of Research and Education through the grant Idei nr. 270, by the French-Romanian
collaboration IN2P3-IFIN and by the ANR NExEN contract.
The calculations have mostly been performed on the GRIF cluster (http://www.grif.fr).

\label{lastpage-01}

\end{document}